\documentclass[10pt,twocolumn]{article}
\usepackage[letterpaper,margin=0.72in]{geometry}
\usepackage[T1]{fontenc}
\usepackage[utf8]{inputenc}
\usepackage{lmodern,microtype,graphicx,booktabs,tabularx,array,seqsplit,natbib,url,xcolor,placeins}
\usepackage[hidelinks]{hyperref}
\title{Revision-Consistent Code Context for Agents That Edit Their Environment}
\author{Meng Shi\\Independent Researcher\\\href{https://orcid.org/0009-0004-8190-7903}{ORCID: 0009-0004-8190-7903}}
\date{}
\begin{document}
\maketitle
\begin{abstract}

Editing agents query a repository while simultaneously changing it. Source content, symbols, graph relations, and retrieval state may be maintained by different stores and refreshed at different times. Eventual convergence is therefore insufficient: during an update, a successful query can combine facts that never belonged to one repository state. We call this condition \emph{mixed-revision code context}.

This paper proposes an Agent-visible \emph{complete-revision publication contract} for composite code-intelligence queries. Before a query-visible publication point, a service must wait, report unavailability, or serve a complete old revision; after that point it serves the complete new revision. The contract does not depend on revision integers or one physical transaction. It requires an equivalent shared version identity and publication boundary across every view returned in one response.

We instantiate the method in Codetrip, whose \texttt{brief} operation joins source, symbol, and graph facts. In a controlled comparison on frozen Gin v1.12.0, a live-source composition strategy produced two unsafe responses in 150 update trials (one mixed and one invalid), whereas complete-revision publication produced zero in 150 under the same five edit scenarios. We further implement journaled, idempotent publication and terminate the real MCP process at five durable boundaries. All 300 evaluable recovery trials on frozen Gin and Cobra (after one separately reported infrastructure exclusion) returned complete-new context equivalent to a clean full rebuild, with median restart-to-context latency of 77--131 ms across cells. In a separate rename probe (30 trials per released system), mixed context occurred in 0/30 Codetrip 0.3.0 trials and 30/30 trials for both CodeGraph 1.5.0 and codebase-memory-mcp 0.9.0 before clean-full-equivalent convergence.

The contribution is not a new isolation primitive. It is an Agent-facing safety contract, implementation pattern, and measurement protocol for the boundary where live repository bytes are joined with asynchronously derived facts. Every successful composite response consequently has a testable repository-state meaning, including after interruption.

\end{abstract}

\section{Introduction}

Repository-level coding agents operate in a loop rather than over a static corpus. They search for a symbol, inspect its callers and contracts, edit one or more files, run tools, and immediately query the repository again. Recent work has improved this loop through iterative retrieval, repository graphs, structured localization, and specialized Agent-computer interfaces \citep{zhang2023repocoder,yang2024sweagent,liu2024codexgraph,ouyang2024repograph,chen2025locagent}. These techniques primarily ask whether the returned context is relevant. They generally leave a separate question implicit: which repository revision does the returned context describe?

That question matters because a code-intelligence response is usually composed from multiple fact domains backed by different stores. Its source excerpt may come from the working tree, its symbol location from a lexical index, its caller set from a graph, and its retrieval metadata from another store. After an edit, these views do not necessarily become current simultaneously. A response can consequently contain the renamed function body from revision \texttt{r+1} and a call edge targeting the old name from revision \texttt{r}. Such a response is not merely stale. It is a state that never existed.

\begin{figure*}[t]
\centering
\includegraphics[width=\textwidth]{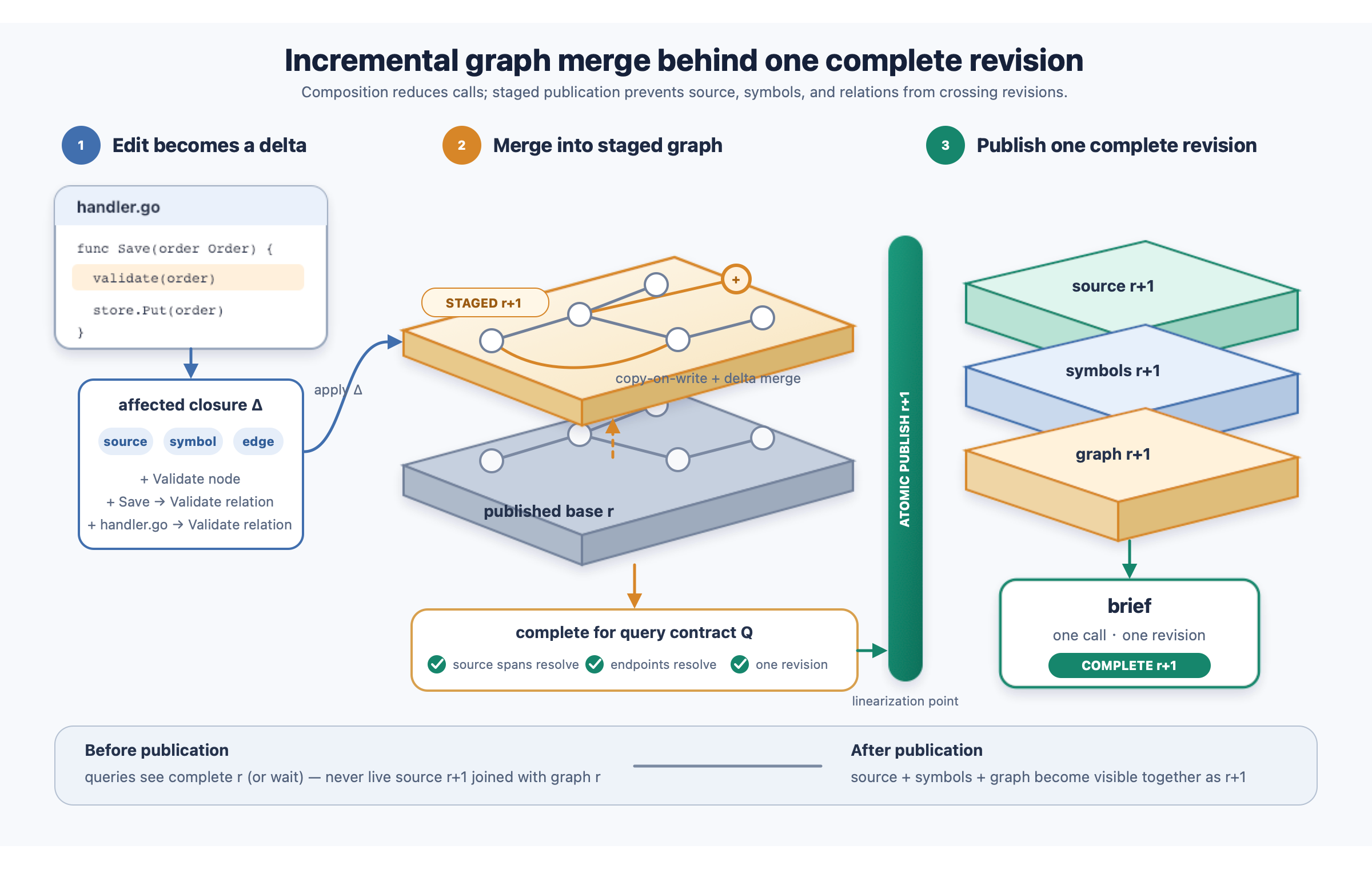}
\caption{Core mechanism and argument. Composition provides interaction efficiency; incremental copy-on-write merging avoids rebuilding unaffected graph facts; and a shared publication boundary provides revision coherence. Before the boundary, queries see complete revision \texttt{r} or wait. After it, all declared views become visible together as complete revision \texttt{r+1}.}
\label{fig:complete_revision_overview}
\end{figure*}

We call this phenomenon \emph{mixed-revision code context}. It differs from ordinary staleness in a way that matters to an Agent. Complete-old context is coherent and can be retried or explicitly labeled stale. Mixed context can look current because some of its most visible fields already reflect the edit. An Agent may then infer a nonexistent broken call site, perform unnecessary recovery reads, or construct a patch from incompatible facts.

The central claim of this paper is that a composite context operation needs a query-visible publication boundary across all views it returns. Before that boundary, the service waits, reports unavailability, or returns a complete old revision; after it, the service returns the complete new revision. New source plus an old graph is not a successful complete-revision response, even when a warning makes the graph's staleness visible.

The revision identifier itself is not the contribution and is not irreplaceable. A transaction, immutable snapshot, MVCC generation, content identity, or atomic manifest may implement the same contract. What cannot be omitted is an equivalent, verifiable boundary that assigns all facts returned by one successful composite query to one complete repository state.

We instantiate the method in Codetrip, an open-source local code-intelligence service that maintains source search, a symbol index, typed graph relations, and bounded Agent context packs. Its \texttt{brief} operation is the concrete composite query used in this study, not the scope of the method itself. To evaluate the effect of a shared revision boundary, we compare two controlled policies under the same workload: unconstrained live-source composition, which may join current workspace source with the last published graph, and complete-revision enforcement, which admits a successful response only when source, symbols, and graph relations belong to one published revision.

We first test whether mixed revisions arise in a real update window. A frozen Gin v1.12.0 workload covers rename, caller addition, signature change, deletion, and move scenarios, with 30 independent repetitions per scenario. All 150 unconstrained-composition trials eventually converge, yet immediate queries produce one mixed response and one invalid response. Eventual convergence is therefore insufficient evidence of query consistency during an update. With complete-revision enforcement, all 150 trials under the same protocol satisfy the query contract, and every final projection matches a clean full rebuild. The Results section reports the complete classification and timing results.

We then test recovery rather than assuming it from the normal execution path. A process-level harness pauses the real MCP server at five durable boundaries, sends \texttt{SIGKILL}, restarts the same data directory, and compares the recovered context with an isolated clean full rebuild. On Gin v1.12.0 and Cobra v1.10.2, all 300 evaluable trials recover complete-new context and match clean full; one additional infrastructure failure is excluded as described under Reproducibility and exclusions.

This paper makes four contributions centered on one claim: composite code context needs a complete-revision publication boundary.

\begin{enumerate}
\item \textbf{Failure model:} it identifies mixed-revision code context as distinct from ordinary staleness and gives it testable observation states.
\item \textbf{Agent-facing contract:} it defines which responses a composite tool may successfully expose, independently of the underlying version mechanism.
\item \textbf{Recoverable realization:} it maps the contract to revision-bound source, a graph authority, staged publication, and journaled roll-forward recovery.
\item \textbf{Boundary-window evidence:} it queries during update, compares controlled publication strategies, checks clean-full equivalence, and kills the real process at durable boundaries. A separate frozen cross-system probe tests whether the failure is confined to Codetrip's former composition strategy.
\end{enumerate}

\section{Motivation}

\subsection{The edit-query loop}

Static retrieval evaluates a query against a fixed corpus. An editing Agent instead executes a stateful sequence:

{\footnotesize
\begin{verbatim}
query(r) -> edit -> durable files(r+1) -> refresh
         -> query(?) -> next action
\end{verbatim}
}

The question mark is the central concern of this paper. During refresh, the working tree is already \texttt{r+1}, while one or more indexes may still represent \texttt{r}. Eventual convergence does not determine what a query is allowed to observe between the edit and convergence.

\subsection{A minimal mixed-revision failure}

Our rename fixture begins with:

{\footnotesize
\begin{verbatim}
func codetripLiveTarget() string {
  return "old"
}
func codetripLiveCaller() string {
  return codetripLiveTarget()
}
\end{verbatim}
}

The edit renames the target and updates its caller:

{\footnotesize
\begin{verbatim}
func codetripLiveTargetRenamed() string {
  return "new"
}
func codetripLiveCaller() string {
  return codetripLiveTargetRenamed()
}
\end{verbatim}
}

A complete-old answer contains the old caller source and an edge to \texttt{codetripLiveTarget}. A complete-new answer contains the renamed caller source and an edge to \texttt{codetripLiveTargetRenamed}. The observed failure combined the new caller source with the old edge. Both fields were individually valid, but their composition described neither repository revision.

\begin{figure*}[t]
\centering
\includegraphics[width=\textwidth]{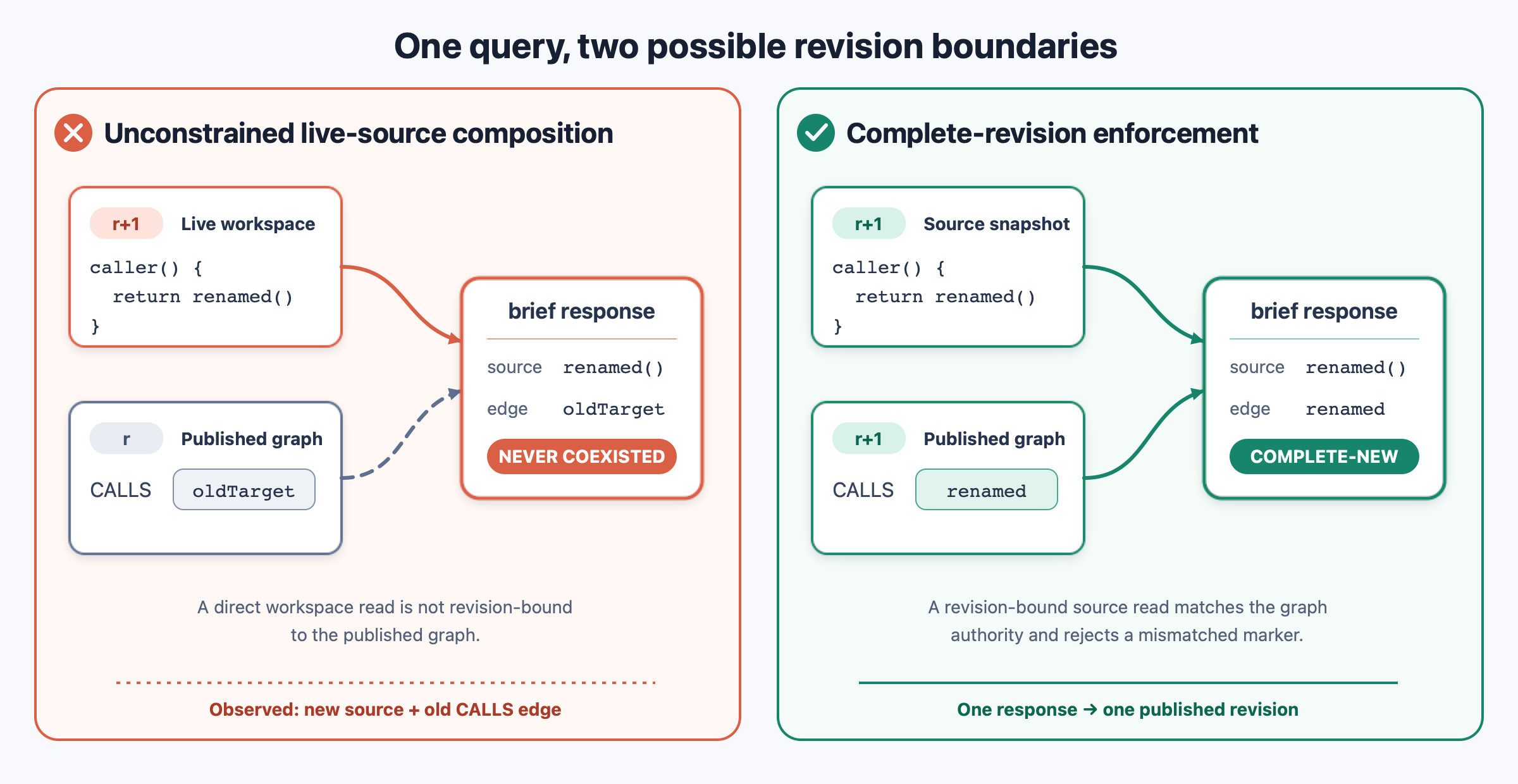}
\caption{Minimal mixed-revision mechanism and the controlled alternative. The left response joins new source with an old call edge; the right response admits the same fields only after they share one published revision.}
\label{fig:mixed_revision_failure}
\end{figure*}

\subsection{Why eventual consistency is insufficient}

If a service explicitly returns complete-old context, an Agent or host can wait for freshness. If it returns unavailable, the host can retry. A warning that the graph is still updating can make staleness visible, but it does not make new workspace source and an old graph belong to one repository revision. Once those fields have been composed, their content alone generally cannot distinguish a cross-revision join from analysis imprecision or a valid program state. The service must therefore preserve and verify the revisions of these views during composition rather than ask the caller to interpret a mixed response.

Nor is successful final convergence sufficient evidence. In our live-source matrix, all 150 trials eventually matched a clean full rebuild, yet two trials violated the query contract immediately after the edit. Measuring only final state or raw refresh duration would have missed both failures.

\subsection{Publication strategies and their guarantees}

An updating code-intelligence service can expose several behaviors. They make different availability choices and provide different guarantees:

\begin{table*}[t]
\centering
\small
\begin{tabularx}{\textwidth}{>{\raggedright\arraybackslash}p{0.19\textwidth}XX}
\toprule
Strategy & Response during graph refresh & Guarantee \\
\midrule
Live-source composition & new workspace source + last published graph & may silently mix revisions \\
Stale-warning composition & new workspace source + old graph + warning & exposes degradation, but remains cross-revision \\
Complete-revision publication & wait, unavailable, complete-old, or complete-new & every successful response belongs to one complete revision \\
\bottomrule
\end{tabularx}
\end{table*}

The strategies exchange availability for meaning. Complete-revision publication makes the successful-response invariant explicit: callers never need to infer whether a source span and a relation endpoint can be combined, because both belong to one named repository state. Section~\ref{sec:threats} discusses the costs and limits of enforcing that invariant.

\section{Model and Contract}

\subsection{Repository revisions and query-visible views}

Let repository state \texttt{S\_r} be an immutable mapping from repository-relative paths to bytes at system revision \texttt{r}. Let \texttt{V\_k(r)} denote the query-visible view of fact domain \texttt{k} derived from \texttt{S\_r}, where \texttt{V} stands for \emph{view}. The fact domains considered in this paper are:

{\footnotesize
\begin{verbatim}
V_source(r)   source excerpts and file identities
V_symbol(r)   symbols, locations, and stable identities
V_graph(r)    typed structural and semantic relations
V_search(r)   lexical or semantic retrieval state
V_meta(r)     analyzer evidence and revision metadata
\end{verbatim}
}

We define views by the kinds of facts exposed in a response, not by CLI commands, code modules, or physical stores. One query may combine several views; for example, \texttt{brief} returns source, symbol, and graph facts. The complete-revision contract applies to every view used and returned by that query: all such views must derive from the same repository revision. Other systems may define additional views as needed.

\subsection{Complete revision and response consistency}

A revision is \emph{complete} for query contract \texttt{Q} when every view required by \texttt{Q} was derived from \texttt{S\_r}, passed its integrity checks, and has resolvable source spans and relation endpoints within \texttt{S\_r}. Completeness does not imply that static analysis perfectly models runtime behavior. It states that the facts the analysis does return share one source state.

For response \texttt{a}, let \texttt{facts(a)} be its query-visible facts and \texttt{rev(f)} the revision from which fact \texttt{f} was derived. Then:

{\footnotesize
\begin{verbatim}
revisions(a) = { rev(f) | f in facts(a) }
\end{verbatim}
}

The response is revision-consistent when \texttt{|revisions(a)| = 1} and that revision is complete for \texttt{Q}. It is mixed-revision when the set contains more than one revision.

\subsection{Observation states}

For a transition from \texttt{r} to \texttt{r+1}, we classify a response as:

\begin{itemize}
\item \textbf{complete-old:} all facts belong to complete revision \texttt{r};
\item \textbf{complete-new:} all facts belong to complete revision \texttt{r+1} and satisfy the edit oracle;
\item \textbf{mixed:} facts combine revisions;
\item \textbf{invalid:} the response violates its schema, refers to missing endpoints, or cannot be assigned a coherent revision;
\item \textbf{unavailable:} the system blocks or explicitly reports that no complete revision is available.
\end{itemize}

Complete-old is a freshness state, not by itself a consistency failure.

\subsection{Complete-revision publication}

A system satisfies complete-revision publication when every successful transition has a query-visible linearization point such that:

\begin{enumerate}
\item queries before the point observe complete \texttt{r} or explicit unavailability;
\item queries after the point observe complete \texttt{r+1};
\item no query observes only a subset of the required new views; and
\item interruption leaves a complete revision or a journaled state that is recovered before queries are enabled.
\end{enumerate}

This behavioral contract does not require all views to be updated by one physical transaction. A system may write new symbols, source content, and graph relations to separate stores in sequence. Until publication completes, queries must either wait or continue reading the complete old revision. The new revision becomes visible only after all views required by the query are ready.

\section{Codetrip Design}

\subsection{Query surface}

Codetrip builds a local typed code graph and associated search indexes. Its \texttt{brief} operation returns a bounded context pack containing a symbol, signature, source, incoming and outgoing relations, contracts, implementations, entrypoints, and related tests.

The contract applies to every query that composes multiple maintained views; \texttt{brief} is the strongest composite query exercised here, not the contract's only possible query surface. A single-view operation needs revision identity but has no cross-view join to tear.

Composite queries also amplify the consistency problem: one response joins facts that otherwise would have been fetched through separate operations. The join is trustworthy only when those facts share a publication boundary. Complete-revision enforcement prevents the convenience of composition from being purchased with incompatible facts.

\subsection{Incremental update}

After a file change, Codetrip computes an affected semantic closure and stages the resulting source, symbol, search, and graph mutations under revision \texttt{r+1}. These mutations may reach separate stores at different times, but none is query-visible as \texttt{r+1} until every view required by the query contract is ready. An error leaves revision \texttt{r} visible.

\subsection{Recoverable publication protocol}

The incremental coordinator requires \texttt{revision = baseRevision + 1} and checks the graph authority's active revision before preparing a commit. Publication then proceeds as follows:

\begin{enumerate}
\item durably save the complete delta in a repository journal;
\item idempotently apply derived stores, including symbols and source;
\item persist index-state metadata;
\item atomically apply the graph mutation and advance its authoritative revision;
\item clear the journal.
\end{enumerate}

The repository operation lock excludes readers during publication, and the graph authority is the query-visible linearization point. On restart, Codetrip finishes a journaled pre-authority revision or recognizes an already published post-authority revision before accepting queries. Appendix A records the store layout, ownership assumption, and boundary-level recovery protocol.

\subsection{Revision-bound source}

Each persisted source view carries the repository revision it represents. A composite read supplies the graph authority's active revision and succeeds only when the source revision matches it. Thus, while workspace files have advanced to \texttt{r+1} and the published graph remains at \texttt{r}, the query reads revision-bound source rather than live files. Every successful response therefore draws its source and derived relations from one complete published revision. Appendix A describes the immutable layers, deletion representation, and replay details.

\section{Evaluation}

\subsection{Research questions}

We ask:

\begin{itemize}
\item \textbf{RQ1 -- Strategy effect:} Compared with unconstrained live-source composition, does complete-revision enforcement prevent mixed or invalid context during the same structural-update workload?
\item \textbf{RQ2 -- Correctness and latency:} Does the constrained strategy preserve clean-full equivalence, and what edit-to-complete-new latency is observed under the protocol?
\item \textbf{RQ3 -- Recovery:} Does the publication contract survive process termination at each durable boundary, and how long does restart-to-context recovery take?
\item \textbf{RQ4 -- Cross-system reachability:} Under one frozen rename fixture, do two independently developed code-context systems expose at least one mixed response before reaching their documented fresh state?
\end{itemize}

\subsection{Normal-update protocol}

The frozen \texttt{paper-draft-v2} protocol uses isolated cold stores and disposable copies of Gin v1.12.0 at commit \texttt{\seqsplit{73726dc606796a025971fe451f0aa6f1b9b847f6}}. Five deterministic edit fixtures cover rename, add-caller, a composite signature change, delete, and move. Each fixture is a predefined edit scenario with old/new source markers, relation-node identity oracles, and absence conditions where applicable.

Each cell contains 30 repetitions. The clock origin \texttt{t\_e} is recorded after all fixture files have been written and durably synchronized. Queries are scheduled at 0, 25, 50, and 100 ms, then every 100 ms until complete-new, followed by ten post-publication observations. Every trial builds an isolated clean full index from the final bytes and compares a canonical projection of source, symbol, and relations. We retain mixed, invalid, unavailable, stale-after-publication, and timeout states separately.

The unconstrained live-source composition policy uses artifact commit \texttt{\seqsplit{a3cbf5c10e4672ddc33c940917c4ce58e6c100a1}}, binary SHA-256 \texttt{\seqsplit{36dcd59977d47679be2e233ab90a1bdba42c30147d48fdc7d9101f9d1a831701}}. The complete-revision enforcement policy uses the same repository, edit fixtures, protocol, and repetition count. The controlled factor is whether a response may compose live workspace source with the last published graph, or must bind every returned fact to the graph's published revision.

\subsection{Process-kill protocol}

The recovery harness starts the real watch-enabled MCP process and verifies a complete-old answer. It durably writes the rename fixture and pauses the process at one of five boundaries:

\begin{enumerate}
\item after journal save;
\item after symbol publication;
\item after source publication;
\item after index-state publication; or
\item after graph-authority publication but before journal clear.
\end{enumerate}

After an atomically written ready signal, the harness sends \texttt{SIGKILL}, restarts the same store, requests \texttt{brief}, and builds an independent clean full store. The trial passes only if the killed process exits by signal, recovery returns complete-new, and the canonical response equals clean full.

We run 30 repetitions per boundary on Gin v1.12.0 and Cobra v1.10.2 at commit \texttt{\seqsplit{88b30ab89da2d0d0abb153818746c5a2d30eccec}}, for 300 evaluable trials. The binary SHA-256 is \texttt{\seqsplit{1fa3acffcb688c7d8e24bf84cdd3c7378d7593d64af8c826774b408759b9cc95}}. Deterministic unit and integration tests and synthetic smoke runs are not pooled with these observations.

\subsection{Cross-system boundary probe}

We run the same Gin rename fixture for 30 isolated trials per released system with Codetrip 0.3.0, CodeGraph 1.5.0~\citep{mchenry2026codegraph}, and codebase-memory-mcp 0.9.0. Every system begins from a clean old index, applies the same durable source edit, classifies raw returned source and call-target facts, and verifies the final result against that product's clean full index.

The product workflows differ. CodeGraph is polled during its documented automatic refresh. codebase-memory-mcp does not automatically refresh this project in the tested workflow, so the probe classifies one immediate query before invoking its documented manual reindex operation. We therefore compare the trial-level outcome--whether any automatic or pre-refresh response is mixed--rather than treating observation counts or refresh latency as a product ranking. Versions and executable hashes are frozen in the aggregate.

\subsection{Reproducibility and exclusions}

Machine-readable aggregates record repository commits, binary and artifact hashes, cell counts, timing summaries, and exclusions. One additional Gin attempt reached its injected boundary and recovered, but the subsequent independent clean-full process failed when the host filesystem had only 24 MiB available; Pebble panicked while rotating its WAL. Because equivalence could not be evaluated, the attempt was classified as an infrastructure failure, retained outside the aggregate, and rerun after enabling bounded per-round cleanup. It is not counted among the 300 evaluable trials.

\section{Results}

\subsection{RQ1: complete-revision enforcement prevents unsafe composition}

The next table reports the normal-update matrix.

\begin{table*}[t]
\centering
\small
\begin{tabularx}{\textwidth}{>{\raggedright\arraybackslash}p{0.19\textwidth}XXXXXX}
\toprule
Build & Trials & Protocol pass & Mixed & Invalid & Timeout & Clean-full equivalent \\
\midrule
Unconstrained live-source composition & 150 & 148 & 1 & 1 & 0 & 150 \\
Complete-revision enforcement & 150 & 150 & 0 & 0 & 0 & 150 \\
\bottomrule
\end{tabularx}
\end{table*}

The mixed rename response occurred 0.764 ms after \texttt{t\_e} and was followed by complete-new at 747.703 ms. Its new workspace source was composed with the old published \texttt{CALLS} relation. The invalid delete response occurred at 0.643 ms and was followed by complete-new at 778.224 ms. Thus, the cross-revision window was short but observable under an immediate Agent query. Final-state testing alone would classify all 150 baseline trials as successful.

The result confirms the mechanism under test: final convergence did not prevent live-source composition from exposing a response with no coherent repository revision. Complete-revision enforcement removes that response class by making coherence a condition of successful publication.

\begin{figure*}[t]
\centering
\includegraphics[width=\textwidth]{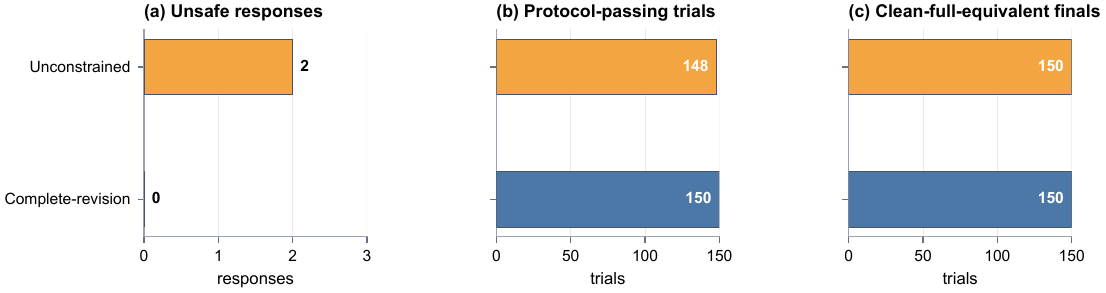}
\caption{Controlled publication-strategy comparison on the same five Gin edit classes. Complete-revision enforcement removes the two observed unsafe responses without changing final clean-full equivalence. The figure is generated directly from the checked-in aggregate JSON files.}
\label{fig:strategy_ablation_bars}
\end{figure*}

\subsection{RQ2: the constrained strategy preserves clean-full equivalence}

All 150 complete-revision trials passed the protocol. There were no mixed, invalid, unavailable, stale-after-publication, or timeout observations, and every final incremental response matched its clean full projection.

Relative to live-source composition, observed unsafe responses decreased from 2/150 to 0/150, while every final result in both strategies remained equivalent to clean full. Each of the five edit classes independently passed 30/30 complete-revision trials.

\begin{table*}[t]
\centering
\small
\begin{tabular}{lrrrrr}
\toprule
Edit scenario & Pass & P50 edit-to-complete-new & P95 & Mixed & Invalid \\
\midrule
Rename & 30/30 & 749.318 ms & 787.889 ms & 0 & 0 \\
Add caller & 30/30 & 1194.010 ms & 1250.726 ms & 0 & 0 \\
Signature change & 30/30 & 775.421 ms & 820.465 ms & 0 & 0 \\
Delete & 30/30 & 787.819 ms & 814.113 ms & 0 & 0 \\
Move & 30/30 & 789.775 ms & 830.748 ms & 0 & 0 \\
\bottomrule
\end{tabular}
\end{table*}

\begin{figure*}[t]
\centering
\includegraphics[width=\textwidth]{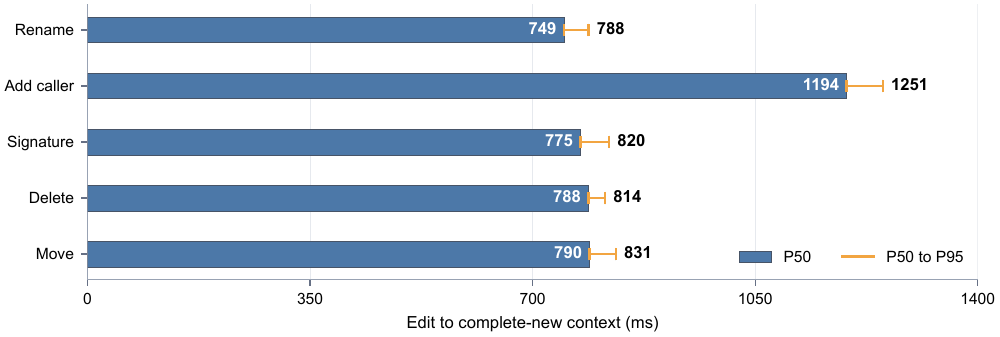}
\caption{End-to-end edit-to-complete-new latency under complete-revision publication. Bars end at P50; thin extensions terminate at P95.}
\label{fig:edit_latency_bars}
\end{figure*}

The signature fixture also adds a helper call so that the external relation oracle can distinguish revisions; its latency must not be interpreted as the cost of a pure signature change. Delete additionally passed its old-symbol absence check in all trials.

The controlled comparison isolates the strategy effect in this matrix: both policies eventually produced the same clean-full-equivalent final state, but only complete-revision enforcement constrained every successful intermediate response to one published revision.

The latency table measures the end-to-end time from durable edit completion to a complete-new response.

\subsection{RQ3: interrupted revisions recovered completely}

Across the two repositories, all 300 evaluable process-kill trials recovered a complete-new response identical to clean full. Here, \emph{evaluable} excludes the one infrastructure failure described under Reproducibility and exclusions.

This is 30/30 successful recoveries in each of ten repository-by-boundary cells. The five interruption points exercise two different obligations: before authority publication, replay must finish the pending revision; after authority publication, recovery must recognize the already visible revision without duplicating derived updates. No evaluated recovery returned a mixed, invalid, or clean-full-mismatching result.

\begin{table*}[t]
\centering
\small
\begin{tabular}{lrrrr}
\toprule
Repository & Boundaries & Trials & Complete-new & Clean-full equivalent \\
\midrule
Gin v1.12.0 & 5 & 150 & 150 & 150 \\
Cobra v1.10.2 & 5 & 150 & 150 & 150 \\
\textbf{Total} & \textbf{10 cells} & \textbf{300} & \textbf{300} & \textbf{300} \\
\bottomrule
\end{tabular}
\end{table*}

Restart latency depends on how much journaled work must be replayed. The post-authority boundary is fastest because recovery only needs to recognize the already published revision and clear the journal.

\begin{table*}[t]
\centering
\small
\begin{tabular}{lrr}
\toprule
Boundary & Gin P50 / P95 & Cobra P50 / P95 \\
\midrule
After journal save & 131.031 / 148.894 ms & 109.120 / 140.667 ms \\
After symbols & 128.620 / 135.956 ms & 107.879 / 119.540 ms \\
After source & 112.374 / 185.272 ms & 93.226 / 102.297 ms \\
After index state & 115.733 / 125.304 ms & 90.860 / 99.825 ms \\
After authority & 86.463 / 95.344 ms & 77.124 / 87.293 ms \\
\bottomrule
\end{tabular}
\end{table*}

\begin{figure*}[t]
\centering
\includegraphics[width=\textwidth]{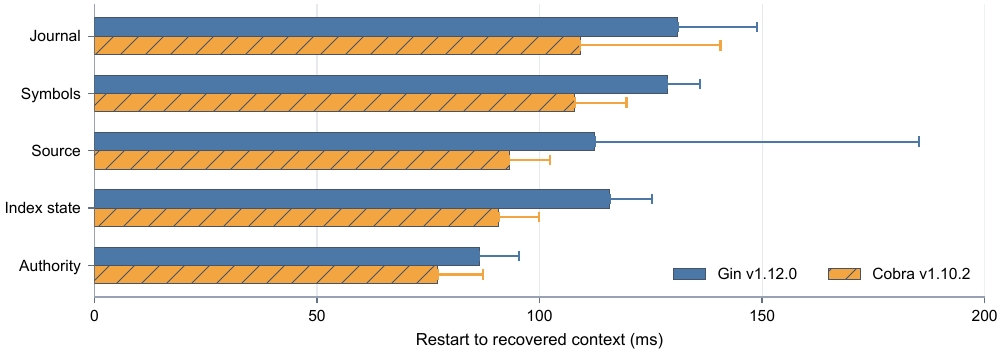}
\caption{Same-store recovery latency by injected process-kill boundary. Bars end at P50; thin extensions terminate at P95. All 300 evaluable trials recovered correctly and matched a clean full rebuild.}
\label{fig:recovery_latency_bars}
\end{figure*}

\subsection{RQ4: mixed context is observable outside Codetrip}

The two comparison systems exposed mixed context before convergence, whereas Codetrip 0.3.0 did not: Codetrip was 0/30, CodeGraph was 30/30 during automatic refresh, and codebase-memory-mcp was 30/30 for immediate queries before its documented manual reindex. Every system subsequently reached a clean-full-equivalent final state.

\begin{table*}[t]
\centering
\small
\begin{tabularx}{\textwidth}{>{\raggedright\arraybackslash}p{0.19\textwidth}XX}
\toprule
System and workflow & Trials with mixed & Fresh and clean-full equivalent \\
\midrule
Codetrip 0.3.0, complete-revision automatic refresh & 0/30 & 30/30 \\
CodeGraph 1.5.0, automatic & 30/30 & 30/30 \\
codebase-memory-mcp 0.9.0, before manual reindex & 30/30 & 30/30 \\
\bottomrule
\end{tabularx}
\end{table*}

\begin{figure*}[t]
\centering
\includegraphics[width=\textwidth]{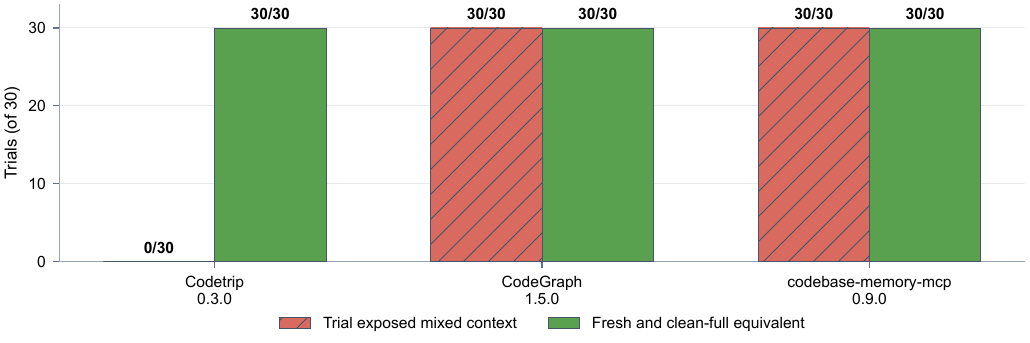}
\caption{Mixed-context reachability under one frozen Gin rename fixture. The hatched bars are the primary update-window outcome; the solid bars show why final convergence alone does not detect the failure class.}
\label{fig:cross_system_bars}
\end{figure*}

The comparison shows that mixed-revision composition is observable outside Codetrip, while final convergence alone cannot distinguish the systems: all three eventually reached a correct final state. For codebase-memory-mcp, the mechanism is persistent pre-refresh staleness rather than an automatic refresh race; nevertheless, composing its old call facts with edited workspace source yields the same Agent-visible mixed-revision state. The result is a reachability comparison for one Go fixture, not a ranking of product quality or natural failure incidence.

\section{Discussion}

\subsection{What the contract buys}

The primary gain is that each successful composite response has one testable repository-state interpretation. A silent semantic ambiguity becomes an explicit systems choice: serve complete-old, wait, report unavailability, or publish complete-new. The Agent no longer spends actions re-reading source and relations merely to determine whether its evidence coexisted, and a host can retry unavailability instead of acting on an impossible repository state.

The contract protects the utility of composite context. \texttt{brief} joins source, symbol, and relation facts in one response; without a shared publication boundary, that convenience may introduce an unmarked cross-revision join.

\subsection{Revision consistency is distinct from analysis precision}

A revision-consistent graph can still omit a dynamic call, and a precise analysis can still be combined with source from another revision. The contract in this paper concerns temporal provenance, not semantic completeness. This separation lets systems state a narrow guarantee without claiming perfect program understanding.

\subsection{The contract does not mandate blocking or revision numbers}

Codetrip serializes repository queries with publication. Other systems may choose to serve complete-old context, return a retryable unavailable state, or use multi-version snapshots. The contract permits all three. It rules out cross-revision composition as a normal complete response. Revision integers are likewise one implementation choice; transactions, generations, manifests, or content identities can establish the same boundary. Which permitted behavior and identity mechanism is best depends on update latency and the host Agent's ability to reason about freshness.

\subsection{Composite context tools need stronger provenance}

A one-call context pack can reduce repeated exploration, but it also creates a join across source and derived facts. The more useful the composite response, the larger the potential silent failure surface. Tool schemas should therefore include or enforce revision provenance rather than treating it as internal metadata.

\subsection{Recovery is part of the query contract}

Atomic rename of a manifest is not by itself a complete recovery argument. Derived stores may become durable before the authority, and the process may terminate after authority publication but before cleanup. The journal must identify the exact pending revision, derived-store updates must be idempotent, and repository opening must complete recovery before accepting queries.

\subsection{Implications beyond Codetrip}

The same boundary appears whenever an Agent-facing service joins live files with asynchronously derived state: language-server caches, repository graphs, vector indexes, test-impact databases, security findings, or build metadata. The implementation need not copy Codetrip's graph authority. It needs an observable contract that can assign the returned facts to one complete state.

\section{Threats to Validity}
\label{sec:threats}

\textbf{Repository and language scope.} The normal-update matrix uses Gin only. The recovery matrix adds Cobra, but both repositories are Go projects and use the same rename/call fixture. Results may differ for other languages, repository sizes, edit classes, filesystems, and operating systems.

\textbf{Observation schedule.} Immediate scheduled queries intentionally increase the chance of observing a short boundary violation. Counts such as 1/150 are not estimates of production incidence. They demonstrate reachability under a declared schedule.

\textbf{Oracle scope.} Clean-full equivalence compares the query projection exposed by the fixture, not every byte in every internal index. Static-analysis errors shared by incremental and clean-full builds would not be detected by this comparison.

\textbf{Implementation and evaluator coupling.} Codetrip's authors designed both the method and harness. Machine-readable fixtures, hashes, raw reports, controlled strategy results, and the disclosed infrastructure failure reduce but do not remove this bias. Independent reproduction is needed.

\textbf{Fault model.} \texttt{SIGKILL} at deterministic durable boundaries tests important crash states but not arbitrary hardware failure, torn filesystem writes, memory corruption, or power-loss semantics on other filesystems.

\textbf{Barrier cost.} The reported edit-to-complete-new latency includes file-watch scheduling, parsing, analysis, indexing, publication, and any query waiting. It does not isolate the marginal latency introduced by the complete-revision barrier. The current evidence therefore supports a consistency benefit and an end-to-end latency observation, not a quantified cost-benefit ratio.

\textbf{Comparative scope.} The cross-system probe covers one Go rename fixture and product-specific documented refresh workflows. It establishes observed reachability in the pinned versions, not natural incidence, retrieval quality, or latency superiority. In particular, codebase-memory-mcp requires manual reindexing in this workflow, so its refresh latency is not compared with an automatic watcher. Its mixed result characterizes the Agent-visible composition of edited workspace source with a deliberately not-yet-refreshed index; it is not evidence that the product violates a claimed automatic-refresh guarantee.

\textbf{Literature coverage.} Our screened set includes repository retrieval, graph-guided agents, incremental analysis, and the closest multi-index systems, but industrial snapshot mechanisms or differently named IDE consistency work may remain undocumented or undiscovered.

\section{Related Work}

\subsection{Repository retrieval and coding agents}

SWE-bench established real-repository issue resolution as an evaluation setting \citep{jimenez2023swebench}, while SWE-agent demonstrated that tool interfaces materially affect Agent behavior \citep{yang2024sweagent}. RepoCoder, AutoCodeRover, Agentless, and SWE-Explore improve repository localization or exploration \citep{zhang2023repocoder,zhang2024autocoderover,xia2024agentless,zhang2026sweexplore}. RepoBench and CrossCodeEval evaluate repository-level and cross-file completion, while ContextBench evaluates context retrieval for coding Agents \citep{liu2023repobench,ding2023crosscodeeval,li2026contextbench}. Repository Memory studies persistent repository knowledge for code localization \citep{wang2025repositorymemory}. Our work does not propose another relevance objective. It focuses on the temporal coherence of the context returned while the repository is changing.

\subsection{Graph-guided context}

CodexGraph, RepoGraph, LocAgent, graph-based CodeRAG, and the Code Graph Model use repository structure to improve context or reasoning \citep{liu2024codexgraph,ouyang2024repograph,chen2025locagent,li2025coderagbigraph,tao2025codegraphmodel}. ARISE exposes a multi-granular program graph through an Agent tool API \citep{seddik2026arise}. DRACO, GraphCoder, RANGER, and DyRetriever likewise use dataflow, code graphs, or partial dependency graphs to retrieve repository context \citep{cheng2024draco,liu2024graphcoder,shah2025ranger,liu2026dyretriever}. These systems establish that graph context is useful; graph use alone is therefore not our novelty claim.

\subsection{Incremental analysis}

Incremental CodeQL, IncA, and prior incremental analysis frameworks study reuse, precision, and update cost after source changes \citep{szabo2023incrementalcodeql,erdweg2026inca,zhou2024incrementalapa,szabo2021incremental}. Reviser specifically studies efficient updates to IDE/IFDS-based data-flow analyses \citep{arzt2014reviser}. This paper adopts the need for incremental recomputation but shifts the observable unit from one analysis result to a composite Agent query spanning source and derived stores.

\subsection{Closest systems}

CodePlan incrementally maintains dependency relations while planning repository edits \citep{bairi2023codeplan}. Its goal includes retaining temporal relations useful for later edit planning, rather than exposing only the current complete repository revision. Code Isn't Memory incrementally maintains lexical, vector, and call-graph indexes and marks stale paths \citep{bhola2026codeisntmemory}. Our distinction is the shared revision and publication contract across the facts returned in one query. SpecAgent leaves incremental re-indexing as future work \citep{ma2026specagent}. These neighbors rule out broad claims that Codetrip is the first incremental graph or multi-index Agent system.

\subsection{Established mechanisms, new system boundary}

Linearization, snapshots, and journals are established storage mechanisms; we do not claim a new database isolation level. Database transactions ordinarily define correctness inside one storage boundary. An Agent tool can still tear a response after those internal transactions complete by joining a live working tree with independently maintained stores. This paper identifies that external composition boundary, defines its successful-response contract, and evaluates the update and recovery windows that final-state benchmarks omit.

\section{Funding and Competing Interests}

This work was conducted independently, without institutional funding or the use of proprietary code, data, equipment, or non-public information. The author declares no competing interests. Codetrip was developed using personal time and equipment as an independent open-source project.

\section{Artifact Availability}

Codetrip is available under the MIT License at <https://github.com/mengshi02/codetrip>. The repository contains the complete-revision and recovery implementation and the tests used to compare the two publication strategies. The \texttt{benchmarks/live\_context} directory contains the durable edit fixtures, normal-update runner, process-kill harness, classification protocol, and checked-in machine-readable aggregates used in this paper. Each aggregate records the frozen subject commit, protocol parameters, trial counts, timing summaries, artifact hashes, and exclusions. The cross-system aggregate additionally records pinned executable hashes and trial-level mixed-response reachability for the three reported systems.

The eight paper figures are stored as SVG files under \texttt{paper/figures}. The four main-text result charts are regenerated from checked-in aggregates by \texttt{paper/figures/generate\_bar\_figures.py}; protocol and harness diagrams are placed in Appendix A. The experiment protocol and fault injection plan are included under \texttt{paper}, together with the claims register, literature-search log, and bibliography. For archival evaluation, the final artifact release will pin the paper commit and provide a single command that checks prerequisites, verifies hashes, and reruns the selected matrix.

\section{Conclusion}

The core problem studied in this paper is the absence of a query-visible publication boundary when a single Agent context operation joins asynchronously maintained code facts. We defined a complete-revision contract: every successful composite response must belong to one complete repository state, while implementations remain free to wait, report unavailability, retain complete-old snapshots, or use any equivalent version-identity mechanism.

Codetrip's \texttt{brief} operation provides one concrete instantiation. Observed unsafe responses decreased from 2/150 with live-source composition to 0/150 with complete-revision publication under the same controlled protocol. Journaled roll-forward recovery then preserved complete-new, clean-full-equivalent context in 300/300 process-kill trials across two frozen Go repositories. A separate three-system rename probe observed mixed responses in 0/30 Codetrip 0.3.0 trials and 30/30 trials for both CodeGraph and codebase-memory-mcp before manual reindex; every system still reached a correct final state.

The broader systems lesson is that relevance and eventual convergence do not give a composite code response a coherent state meaning. Editing-agent infrastructure needs a shared, testable publication boundary across every view returned together. Multi-language update evaluation, isolated barrier-cost measurement, context-pack compression, and edit-then-query Agent experiments remain necessary next steps.

\appendix

\section{Implementation and Harness Details}

\subsection{Durable source representation}

Incremental source publication uses immutable layers and deletion tombstones. Because publishing a layer and advancing its revision marker are separately crashable, replay recognizes an already durable layer and repairs its marker without duplicating the layer. A migration initializes a missing legacy marker only when source and graph share the same physical snapshot identity; it never overwrites an existing mismatch.

Codetrip assumes one service process owns a data directory. Its repository operation lock is in-process, while the underlying database lock rejects a second process opening the same store. Concurrent writers sharing one data directory are outside the supported execution model.

\begin{figure*}[t]
\centering
\includegraphics[width=\textwidth]{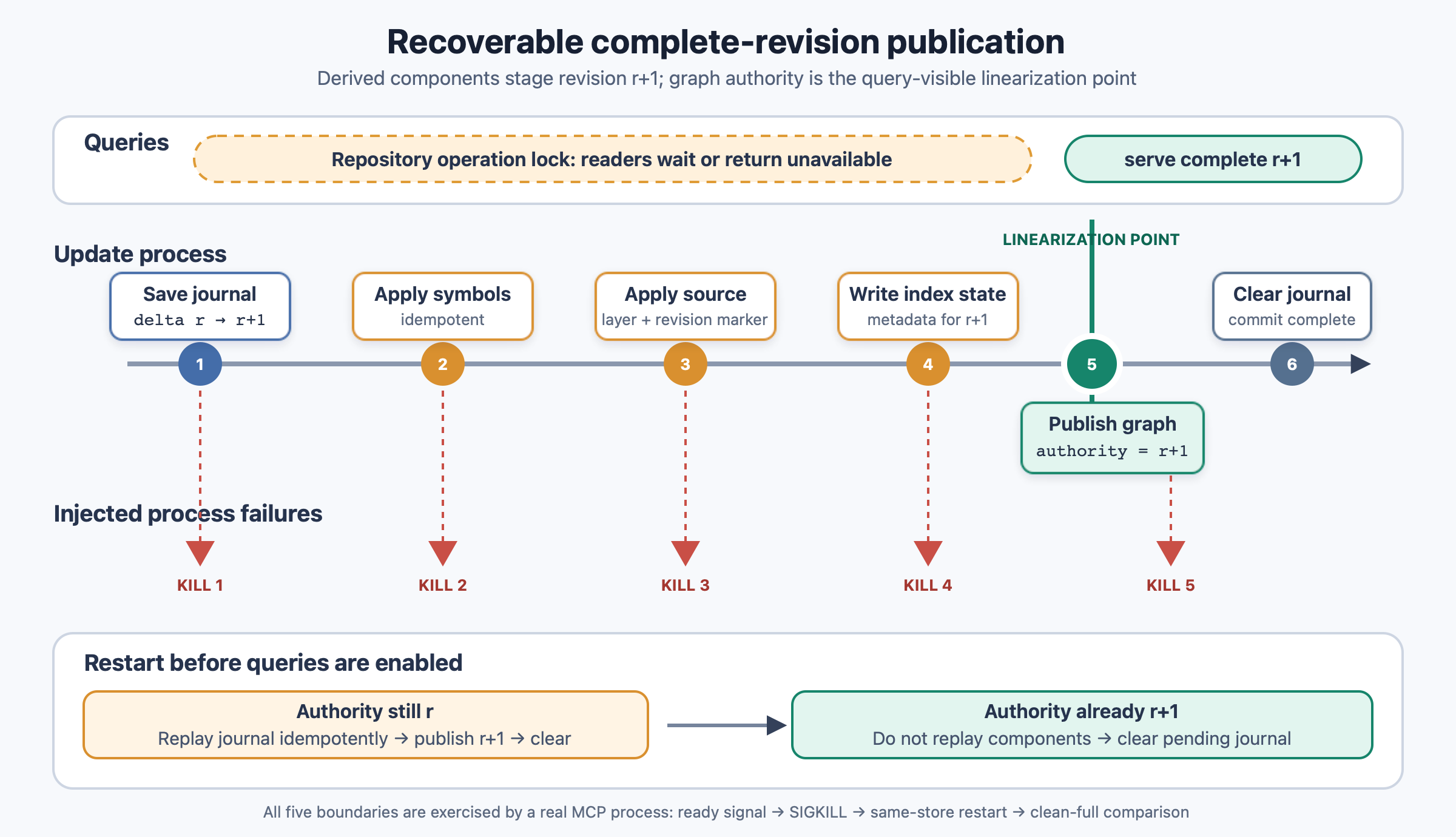}
\caption{Boundary-level publication and recovery protocol. Numbered red markers denote the five experimental \texttt{SIGKILL} boundaries.}
\label{fig:publication_recovery_protocol}
\end{figure*}

\subsection{Process-level fault harness}

\begin{figure*}[t]
\centering
\includegraphics[width=\textwidth]{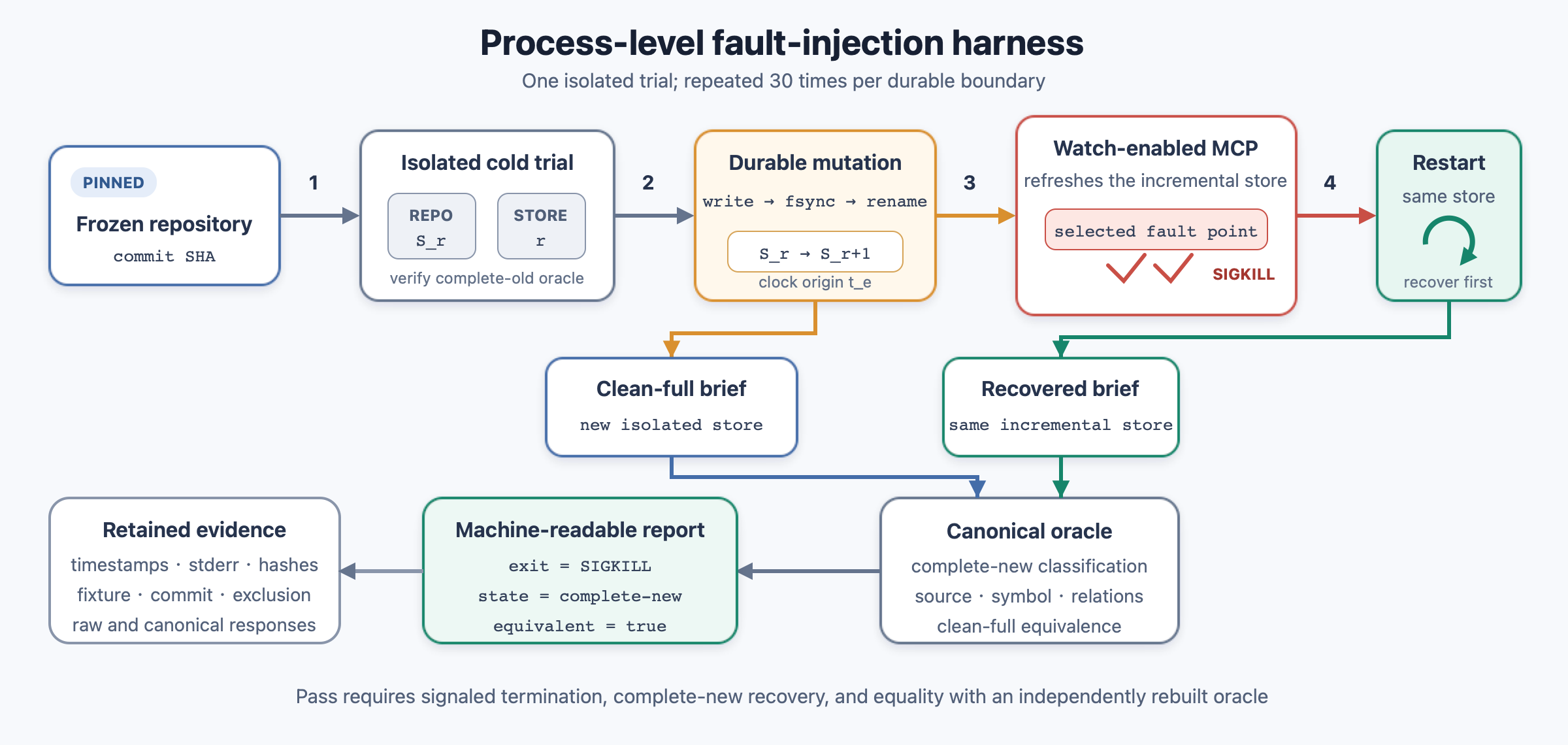}
\caption{One process-kill trial. Each durable boundary and repository is repeated 30 times; passing requires a signaled exit, complete-new recovery, and clean-full equivalence.}
\label{fig:fault_injection_harness}
\end{figure*}
\FloatBarrier
\bibliographystyle{plainnat}
\bibliography{references}
\end{document}